\documentstyle[11pt,aaspp]{article}
\newcommand{\postscript}[2]
 {\setlength{\epsfxsize}{#2\hsize}
  \centerline{\epsfbox{#1}}}
\def\ref#1{\par\noindent \hangindent=0.4in \hangafter=1 #1 \par}
\def\eqalign#1{\null\,\vcenter{\openup\jot \m@th
  \ialign{\strut\hfill$\displaystyle{##}$&$
     \displaystyle{{}##}$\hfill \crcr#1\crcr}}\,}
\def\tempest%
{\begin{array}{ccc}
1 & 1 & 1 \\
1 & 1 & 1 \\
4 & 3 & 8
\end{array}}

\begin{document}

\title{PIXEL LENSING EXPERIMENT TOWARD THE M31 BULGE${}^{1}$}

\author{Cheongho Han}
\affil{Ohio State University, Department of Astronomy, Columbus, OH 43210}
\affil{cheongho@payne.mps.ohio-state.edu}
\footnotetext[1]{submitted to {\it Astrophysical Journal}, Preprint:
OSU-TA-23/95}
 
\begin{abstract}

I demonstrate that one can detect pixel gravitational microlensing 
events at the rate $\sim 180\ {\rm events}\ {\rm yr}^{-1}$ and 
that some fraction of events ($\sim 15$) will be good enough to measure 
time scales at the $20\%$ level if the experiment is carried out 
toward the central M31 bulge region with a dedicated 1 m telescope 
and a $2{\rm K}\times2{\rm K}$ CCD.
I also show that the event rate could be significantly increased (up 
to a factor 6.5) if more aggressive observation strategies are adopted.
These aggressive strategies include multiple observation sites, 
using a bigger telescope,
and reduction of seeing differences among images.
A few years of observation could provide
a data set big enough to extract important information 
about the M31 bulge such as the optical depth distribution, the mass 
spectrum of M31 MACHOs, and the luminosity function,
which currently cannot be obtained by other methods.
Additionally, I find that the extinction of M31 bulge light by 
dust in the disk is {\it not} important and can be approximated 
by the model in which only $20\%$ of light in $V$ band is extinguished 
over most of the bulge area.

\end{abstract}


\newpage
\section{Introduction}

   Pixel lensing, gravitational microlensing of unresolved stars, 
was first proposed by Crotts (1992) and Baillon et al.\ (1993).
Classical lensing experiments are restricted to monitoring only 
fields of resolvable stars, e.g.,\ toward the Galactic bulge and 
Magellanic Clouds (Alcock et al.\ 1995; Udalski et al.\ 1995; 
Ansari et al.\ 1995; Alard et al.\ 1995). 
By contrast, pixel lensing can overcome this restriction and thus 
can extend our ability to probe both Galactic and extragalactic 
Massive Compact Objects (MACHOs). 
Currently, both groups have demonstrated the technique by detecting 
various types of variable stars (Ansari et al.\ 1996; 
Tomaney \& Crotts 1996)

For classical lensing, the light curve is obtained by subtracting 
the reference flux, $B$, from the amplified flux, $F$: 
$$
F_i (A-1) = F -  B;\ \ B = F_i + B_{\rm res},
\eqno(1.1)
$$
where $F_i$ is the flux of the lensed star after (or before) 
being lensed and $B_{\rm res} = \sum_{j\neq i} F_j$ is 
the residual flux from stars not participating in the event.
The amplification is related to the lensing parameters by
$$
A(x) = {x^2 +2 \over x(x^2 + 4)^{1/2} };\ \
x^2 = \omega^2 (t-t_0)^2 + \beta^2,
\eqno(1.2)
$$
where $\omega = t_{\rm e}^{-1}$ is the inverse Einstein ring crossing 
time, $\beta$ is the angular impact parameter in units of angular 
Einstein ring radius, $\theta_{\rm e}$, and $t_0$ is the time 
of maximum amplification.
The angular Einstein ring radius is related to the physical Einstein 
ring radius, $r_{\rm e}$, and to the physical parameters 
of the MACHO by
$$
\theta_{\rm e} = {r_{\rm e} \over D_{\rm ol} };\ \ 
r_{\rm e} = \left( {4GM \over c^2} {D_{\rm ol}D_{\rm ls}
\over D_{\rm os} } \right)^{1/2},
\eqno(1.3)
$$
where $D_{\rm ol}$, $D_{\rm ls}$, and $D_{\rm os}$ are the distances 
between the observer, source, and lens, and 
$M$ is the mass of the lens.
On the other hand, it is impossible to measure directly the absolute values 
of $F$ and $B$ 
for the case of pixel lensing because stars cannot be resolved.
However, the light curve, $F_i (A-1)$, still can be obtained by subtracting 
the reference image from the image in which a lensing event is occurring.

Due to its characteristic differences from classical microlensing
observations, additional analysis is required to estimate the
pixel lensing event rate.
For classical lensing, an event with modest maximum amplification
(e.g., $A_{\rm max} \sim 2$) can often be detected with high 
signal-to-noise ratio, $S/N$, when observed for a few minutes 
on a 1 m telescope.
This is partly because source stars are nearby, and thus bright, 
and partly because the residual flux, $B_{\rm res}$, is moderate or negligible.
Note, however, that $B_{\rm res}$ is not exactly zero due to the blending 
of star light in dense fields.
For the case of pixel lensing, however, stars are located relatively 
far away, and thus yield low signal.
Additionally, the signal suffers from significant noise from 
background stars that are not being lensed.
Therefore, the event detection rate per observed star is significantly 
lower than that from classical lensing and thus events can be detected 
only when either the amplification or the intrinsic flux is very high.
However, the loss of detection rate per star is compensated by the  
fact that many more stars are monitored.
Therefore, one can obtain an equivalent number of total detectable 
events to that of classical lensing experiments.

In this paper, I show that one can detect pixel lensing events 
at a rate $\Gamma \sim 180\ {\rm events}\ {\rm yr}^{-1}$ with a 
dedicated 1 m telescope and a $2{\rm K}\times{\rm 2K}$ CCD.
Among the events, $\sim 15$ events have high enough $S/N$ to measure 
the time scale at the 20\% level.
The rate can be increased by up to a factor $\sim 7$ by employing a few 
improved observational strategies, and thus 
$\sim 100\ {\rm events}$ with measurable time scale could
be detected per year.
These strategies are discussed in \S\ 7.
Gould (1996) developed a general formalism for the analysis of 
pixel lensing.
The reader is referred to this work for a better understanding of 
the formalism that I use without detailed derivation.
Comparison of the formalism with that developed by Colley (1995) is also 
recommended.

\section{Formalism}

The signal-to-noise ratio during a single observation is 
given by
$$
(S/N)_i^2 = {F^2 \alpha t_{\rm cyc} \over 
\Sigma \Omega_{\rm psf} } [A(t_i)-1]^2, 
\eqno(2.1)
$$
where 
$F$ is the flux of the lensed star, 
$\Sigma$ is the surface brightness of observed area, $\Omega_{\rm psf}$ 
is the angular area of the point spread function (PSF),
$t_{\rm cyc}$ is the observation duty cycle,
$\alpha$ is the mean photon detection rate averaged over the duty cycle. 
If the observations are carried out for 4 hrs every night and the camera can 
detect $\epsilon\ {\rm photons}\ {\rm hr}^{-1}$, then 
$t_{\rm cyc} = 1$ day
and $\alpha = \epsilon\times (t_{\rm exp}/24)\ {\rm photons}\ {\rm hr}^{-1}$, 
where $t_{\rm exp}$ is the mean exposure time per day.
Individual images obtained during the whole event are combined, yielding 
the total signal-to-noise ratio 
$$
(S/N)^2 = {F^2 \alpha \over  \Sigma \Omega_{\rm psf}} t_{\rm cyc} 
\sum_i [A(t_i)-1]^2.
\eqno(2.2)
$$
By converting the sum
$t_{\rm cyc}\sum_i [A(t_i)-1]^2$ into an integral 
$\int_{-\infty}^{\infty} dt [A(t)-1]^2 $, 
one can obtain a compact form of $S/N$,
$$
(S/N)^2 = { \pi F^2 \alpha \over 
\Sigma \Omega_{\rm psf} \omega } {\zeta (\beta) \over \beta};\ \ 
\zeta (\beta) = {\int_{-\infty}^{\infty} dt [A(t)-1]^2 \over 
\int_{-\infty}^{\infty} dt (\omega^2 t^2 + \beta^2)^{-1}},
\eqno(2.3)
$$
where the factor $\zeta$ is evaluated and shown in Figure 1 
of Gould (1996).

Events are detected either when the star is bright or the 
amplification is high.
This can be quantified by setting a maximum impact parameter 
$\beta_{\rm max}$ within which a lensing event with flux $F$ 
is detected for a certain threshold
signal-to-noise ratio $(S/N)_{\rm min}$.
The maximum impact parameter is then found by solving equation 
(2.3), i.e., $\beta_{\rm max}$ is the solution of the equation
$$
{\zeta (\beta_{\rm max}) \over \beta_{\rm max}} =
\left( {\Sigma \Omega_{\rm psf} \bar{\omega} \over 
\pi F^2 \alpha } \right) (S/N)_{\rm min}^2,
\eqno(2.4)
$$
where $\bar{\omega}=\bar{t}_{\rm e}^{-1}$ is the inverse of median 
Einstein time scale.
Once $\beta_{\rm max}$ is determined, the event rate
per angular area $d\Omega$ is computed by
$$
{d\Gamma \over d\Omega } 
= {\cal C}_{\rm n} \sum_j \Gamma_0 \beta_{\rm max} (F_j) \phi (F_j);\ \ 
\Gamma_0  =  {2 \over \pi} \bar{\omega}  \tau,
\eqno(2.5)
$$
where $\tau$ is the optical depth, 
$\phi (F_j)$ is the luminosity function (hereafter LF) of M31 bulge stars, 
and ${\cal C}_{\rm n}$ is a normalization factor, 
${\cal C}_{\rm n} = \Sigma [\sum_j \phi (F_j)F_j ]^{-1}$.

\section{Optical Depth Distribution}

For the computation of the event rate, it is necessary to compute 
the optical depth [see eq.\ (2.5)].
The computation of optical depth toward M31
requires some geometrical transformations
since M31 is highly inclined, $i \sim 77^{\circ}$ (Kent 1989).
I define the intrinsic coordinates, $(x,y,z);\ r^2 = x^2 + y^2$, 
that are centered at the center of M31 and $r$ and $z$ are measured 
along and normal to the disk plane, respectively.
These coordinates are related to another set of coordinates, 
$(x',y',z')$, which have the same center but are measured along 
the projected major, minor axis, and the line-of-sight, by,
$ r = [ (y' \cos i + z' \sin i)^2 + x'^2 ]^{1/2},\ 
z = | y' \sin i - z' \cos i |.  $
The adopted distance to M31 is $d_{\rm M31}=770\ {\rm kpc}$.

There are several populations of lenses when stars in the M31 bulge are 
monitored.
The first is MACHOs in the Galactic halo.
The mean optical depth to a single M31 star lensed by the Galactic 
halo MACHOs is
$$
\tau_{\rm MW,halo} = {4\pi G \over c^2}
\int_{0}^{d_{\rm M31}} dD_{\rm ol}\rho_{\rm MW,halo}(D_{\rm ol})D,
\eqno(3.1)
$$
where $D=D_{\rm ol}D_{\rm ls}/D_{\rm os}$.
For the Galactic halo matter distribution, I adopt a
modified isothermal sphere with a core radius $r_{\rm c,MW}$;
$$
\rho_{\rm MW,halo} (r) =
\cases{
\rho_{0} (r_{\rm c,MW}^2 + R_0^2)/(r_{\rm c,MW}^2 + r^2),
&  $r \le 200\ {\rm kpc}$ \cr
0, & $r > 200\ {\rm kpc}$,\cr}
\eqno(3.2)
$$
where the normalization is $\rho_0 = 7.9\times 10^{-3}\ 
M_{\odot}\ {\rm pc}^{-3}$, $R_0=8\ {\rm kpc}$ is the solar 
galactocentric distance, and the core radius is 
$r_{\rm c,MW}=2\ {\rm kpc}$ (Bahcall, Schmidt, \& Soneira 1983).
The value of the Galactic core radius is uncertain, but 
it hardly affects the optical depth determination because M31 
is located $\sim 119^{\circ}$ away from the Galactic center.
With this model, I find $\tau_{\rm MW,halo}=4.4\times 10^{-7}$.

The second lens population is the MACHOs in the M31 halo.
Since the rotation curve of M31 is flat at 
large radii up to $r=30\ {\rm kpc}$ from the center (Kent 1989a), 
the M31 halo is modeled by a standard isothermal 
model with a core radius $r_{\rm c,M31}$;
$$
\rho_{\rm M31,halo}={v^{2}_{\rm c,M31}\over 4\pi G(r^2+r_{\rm c,M31}^{2})},
\eqno(3.3)
$$
where the core radius is determined from the relation
$$
M_{\rm c} = {v_{\rm c,M31}^2\over G}
\int_0^{\infty} drr 
\left(
{1\over r^2}-{1\over r^2+r_{\rm c,M31}^2}
\right).
\eqno(3.4)
$$
With the total core mass of 
$M_{\rm c,M31} = M_{\rm M31,disk}+M_{\rm M31,bulge} = 
13.2\times 10^{10}\ M_{\odot}$ (see below) and the M31 rotation velocity of
$v_{\rm c,M31}=240\ {\rm km\ s}^{-1}$, I find that 
$r_{\rm c,M31} = (2/ \pi)(GM_{\rm c,M31}/
v_{\rm c,M31}^2)=6.5\ {\rm kpc}$.
Then the optical depth contributed by M31 halo lenses is computed 
similarly to equation (3.1), and results in 
$\tau_{\rm M31,halo}=1.91\times 10^{-6}$.

Finally, the rest of the optical depth toward the M31 bulge is 
contributed by the M31 disk+bulge self-lensing events.
This contribution of optical depth is computed by
$$
\tau_{\rm M31} =
{4\pi G \over c^2}
\int_{d_1}^{d_2} dD_{\rm os}\rho_{\rm M31}(D_{\rm os})
\int_{d_1}^{D_{\rm ol}} dD_{\rm ol} \rho_{\rm M31} (D_{\rm ol}) D
\left[ \int_{d_1}^{d_2} dD_{\rm os}\rho_{\rm M31}(D_{\rm os}) \right]^{-1},
\eqno(3.5)
$$
where $\rho_{\rm M31}=\rho_{\rm M31,disk} + \rho_{\rm M31,bulge}$,
$\rho_{\rm M31,disk}$ and $ \rho_{\rm M31,bulge}$
are the M31 disk and bulge densities, and
$d_1=d_{\rm M31} - 30\ {\rm kpc}$ and $d_2=d_{\rm M31} + 30\ {\rm kpc}$
are the lower and upper boundaries of mass distribution
along the line of sight.
The M31 bulge is modeled as an oblate spheroid with its
(unnormalized) matter density as a function of semimajor axis
provided by Kent (1989b).
The best-fitting value of the axis ratio, $c/a=0.75$, is found by fitting
the computed surface brightness (corrected for $I$ extinction)
to a $V$-band image of M31 bulge (see \S\ 5 for the observations).
In the central region where the M31 bulge mass dominates, the
rotation velocity is $\sim 275\ {\rm km}\ {\rm s}^{-1}$.
Thus I normalize the total M31 bulge mass within $r=4$ kpc by
$M_{\rm M31,bulge}=(275\ {\rm km\ s}^{-1}/v_{\rm c,MW})^{2}M_{\rm MW,bulge}$,
where $v_{\rm c,MW}=220\ {\rm km\ s}^{-1}$ and
$M_{\rm MW,bulge}=2\times 10^{10}\ M_{\odot}$ are the adopted rotation speed
of the Galaxy and Galactic bulge mass within $r=4$ kpc
(Zhao, Spergel, \& Rich 1995).
This normalization procedure results in
$M_{\rm M31,bulge}=4\pi\int_{0}^{\infty} dr r^2\rho_{\rm M31,bulge}=
4.9\times 10^{10}\ M_{\odot}$, which is in reasonable agreement with the
determined value of $4.0\times 10^{10}\ M_{\odot}$ by Kent (1989b).
The M31 bulge is cut off at $r\sim 8\ {\rm kpc}$.
The M31 disk is modeled by the double exponential disk, i.e.,
$$
\rho_{\rm M31,disk}(R,z) = {\Sigma_0 \over 2h_z}
\exp\left( -{z\over h_z}\right)
\exp\left( -{R\over h_R }\right),
\eqno(3.6)
$$
where $\Sigma_{0}=280\ M_{\odot}\ {\rm pc}^{-2}$ is the normalization, and
the radial and vertical scale heights are
$h_{z} = 400\ {\rm pc}$ and $h_R = 6.4\ {\rm kpc}$, respectively
(Gilmore et al.\ 1990; Gould 1994).
This results in
$M_{\rm M31,disk}(R\le 30\ {\rm kpc})=
8.3\times 10^{10}\ M_{\odot}$.

The resulting optical depth distribution caused by M31 population
(M31 disk+bulge self-lensing plus M31 halo) is shown as a contour 
map in Figure 1.
The total optical depth both by M31 and Galactic halo lenses are 
obtained by adding $\tau_{\rm MW,halo}=4.4\times 10^{-7}$ to the values marked 
on the contour.
The contour levels are marked on the map in units of $10^{-6}$.
I note that the determined optical depth is subject to
many uncertainties, especially in the mass distributions.
For example, Braun (1991) has argued that there is no need to invoke a
massive dark halo component.
Instead of a halo, his model has higher disk and bulge densities,
c.f., $M_{\rm M31,bulge}= 7.8\times 10^{10}\ M_{\odot}$ and
$M_{\rm M31,disk}= 12.2\times 10^{10}\ M_{\odot}$ within 30 kpc
from the center.
Pixel lensing experiments may resolve the conflicts in the matter density
distribution in M31 from the comparison between the theoretical predictions
and future observational result.


\section{Time-scale Distribution}

Not only the optical depth but also the median time scale 
is required for the computation of the event rate.
The time scale is determined by the combination of both the transverse 
speed, $v$, and the Einstein ring radius of the lens, i.e., 
$t_{\rm e} = r_{\rm e}/v$.
The transverse velocity is related to the lens geometry
and velocity of the source, ${\bf v}_{\rm s}$, the lens, 
${\bf v}_{\rm l}$, and observer, ${\bf v}_{\rm o}$ by
$$
{\bf v} = {\bf v}_{\rm l} - \left[ {\bf v}_{\rm s}
\left( {D_{\rm ol} \over D_{\rm os}} \right) + {\bf v}_{\rm o}
\left( {D_{\rm ls} \over D_{\rm os}} \right) \right].
\eqno(4.1)
$$
For M31-M31 self-lensing events, one can approximate
equation (4.1) as ${\bf v} = {\bf v}_{\rm l} - {\bf v}_{\rm s}$ since
$D_{\rm os} \sim D_{\rm ol}$. 
Then the rms transverse velocity is found by
$\langle v^2 \rangle^{1/2}  = \langle v_{\rm l}^2 + v_{\rm s}^2 -2 
({\bf v_{\rm l} \cdot v_{\rm s}}) \rangle^{1/2} = \sqrt{2}
\langle v_{\rm s}\rangle $ 
because the cross-term 
$\langle {\bf v}_{l} \cdot {\bf v}_{s} \rangle = 0$.
Here $v_{\rm s} \sim \sqrt{2} \sigma_{\rm bulge}$, and 
$\sigma_{\rm bulge} \sim 156\ {\rm km\ s^{-1}}$ 
is the adopted velocity dispersion of the M31 bulge (Lawrie 1983).
The Einstein ring radius is related to the location of the 
lens system by 
$r_{\rm e} \propto D^{1/2}$,
assuming all lenses have the same mass.
In the computation, I assume $M = 0.3\ {\rm M}_{\odot}$ 
because the most probable lens population is low mass stars.
For other masses $t_{\rm e} \propto M^{1/2}$.
The probability of finding an event located 
at a certain location $D$ is found by
$$
P(D) \propto 
\int_{d_1}^{d_2}dD_{\rm os} \rho(D_{\rm os})
\int_{d_1}^{D_{\rm ol}} dD_{\rm ol} \rho(D_{\rm ol})
D^{1/2}.
\eqno(4.2)
$$
Then the time scale probability distribution is 
computed from $P(D)$ by
$$
f(t_{\rm e}) = \int_0^{\infty} dt'_{\rm e} \delta 
\left[ t'_{\rm e} - {(4GMD)^{1/2} \over vc} \right] P(D).
\eqno(4.3)
$$
I find $\bar{t}_{\rm e}$ by taking the median value of $f(t_{\rm e})$.
The resulting median time scale $\bar{t}_{\rm e}$ for M31 
self-lensing events as a function of 
position $(x',y')$ shown in units of days in Figure 2 as a contour map.
For other population events, i.e., Galactic halo and M31 halo events, 
the time scale distributions are similar to that of M31 
self-lensing events (Han \& Gould 1996b).

\section{Extinction by the Dust in the M31 Disk}

The light from the M31 bulge suffers some extinction by dust
in the disk because
half the M31 bulge lies in front of the M31 disk while the other
half lies behind it.
Hence the bulge light from behind the disk is affected by the dust. 
By experiencing extinction, the apparent LF of the stars 
behind the disk will be different from that of stars in front, and thus 
$\Gamma$ is affected by the extinction [see eq.\ (2.5)].
Since the disk is tilted, the total light of stars influenced by the 
extinction differs from position to position.

To quantify the amount of extinction, CCD images in $V$ and $A$
bands were obtained on the night of 1995 August 4 using Tek2K
detector on 1-m United Sates Naval Observatory Ritchey Chr\'etien
Aplanatic telescope,
with a plate scale of $0.\hskip-2pt ''68\ {\rm pixel}^{-1}$.
The $A$ band filter is centered at $6900\AA$ and has a band
width of $\Delta\lambda = 2500\AA$.
Bias and sky subtraction of the images were carried out using
the interactive data reduction package IRAF.

I find that the disk of M31 is quite transparent.
Let $\Sigma_{\rm back}$ and $\Sigma_{\rm front}$ be the surface
brightnesses contributed by stars located behind and in front of
the disk, respectively.
Then the flux ratio between two bands is given by
$$
{\cal R}'(x',y') = { \Sigma_{\rm {\it V},front} + 
f_V \Sigma_{\rm {\it V},back} \over
\Sigma_{\rm {\it A},front} + f_A \Sigma_{\rm {\it A},back} },
\eqno(5.1)
$$
where $f$ is the fraction of light that passes through the disk
in each band.
Let ${\cal R}'_{\rm crit} = \Sigma_{V} / \Sigma_{A}$ be 
the (instrumental) ratio of flux in the two bands in the absence of 
extinction.
This quantity can be determined by measuring the photon counts 
along the far minor axis where nearly all of the bulge is in front of the 
disk and where, consequently, extinction has almost no effect
on the color.
Then, one can rewrite equation (5.1) as
$$
{\cal R} = {{\cal R}'(x',y')\over {\cal R}'_{\rm crit}}  = 
{ \Sigma_{\rm front} + 
f_V \Sigma_{\rm back} \over
\Sigma_{\rm front} + f_A \Sigma_{\rm back} }.
\eqno(5.2)
$$
The fraction $f_{A}$ is related to $f_{V}$ by
$$
f_A = f_{V}^{q};\ \ q = A_A / A_V \sim 0.8,
\eqno(5.3)
$$
which implies more absorption in $V$ band, i.e., ${\cal R} \le 1.0$.
The surface brightnesses, $\Sigma_{\rm back}$ and
$\Sigma_{\rm front}$, are computed from the mass model
in \S\ 3 and the color ratio ${\cal R}(x',y')$
is obtained from the images of the M31 bulge.
Then $f_V$ is determined by solving equation (5.2) numerically
using the relation (5.3),
i.e., by finding the solution of the equation,
$$
f_{V}^{q} - {1 \over {\cal R}} f_V +
{\Sigma_{\rm front} \over \Sigma_{\rm back}}
\left( 1- {1\over {\cal R}} \right) = 0.
\eqno(5.4)
$$
In general, equation (5.4) has two solutions, e.g., for 
${\cal R} = 1$ both $f_{V}=0$ (totally opaque) and  
$f_{V}=1$ (totally transparent) solutions are possible.
For the actual data, one set of solutions has $\langle  f_V \rangle \sim 0.1$,
while the other set has $\langle  f_V \rangle \sim 0.8$.
If $f_{V}=0.1$, the M31 bulge surface brightness 
distribution would be extremely asymmetric with respect to the 
photometric major axis because most of the light 
would be extinguished on the near side while only a small 
fraction would be lost on the far side due to the high inclination of M31.
In fact, the isophotes are close to symmetric.
The mean extinction is therefore $f_{V}=0.8$.
Because the disk is optically thin, the overall event rate depends only
on the mean extinction.
I therefore adopt the simplified model of uniform extinction with 
$f_V \sim 0.8$, which corresponds to $f_I \sim 0.88$.

\section{Realistic Event Rate Estimate}

\subsection{Additional Restrictions}

The event rate formula in equation (2.5) is derived under the 
assumption that the dominant form of noise comes from
source stars, i.e., the Poisson photon noise $N_{\gamma}$.
However, except in the central region, 
$\sim (6.5 \times 6.5)\ {\rm arcmin}^2$, 
most of the M31 bulge region has a surface brightness less than
the sky brightness, which is $\langle I \rangle = 19.5\ 
{\rm mag}\ {\rm arcsec}^{-2}$ averaged over the phases of the Moon. 
Therefore, the noise due to the sky-background, $N_{\rm sky}$, 
becomes important.
In addition, the differences in PSF from image to image taken 
at different times cause extra noise (Gould 1996),
which I denote $N_{\rm psf}$.
I define the additional factors in noise due to sky 
and PSF variation by
$$
\eta_{\rm sky} = \left( {N_{\gamma}^2 +
N_{\rm sky}^2 \over N_{\gamma}^2 } \right)^{1/2},\ \
\eta_{\rm psf} = { N_{\rm psf} \over N_{\gamma} }.
\eqno(6.1.1)
$$
The values of these factors are computed by
$$
\eta_{\rm sky} = \left( {\Sigma_{\rm M31} + \Sigma_{\rm sky}
\over \Sigma_{\rm M31} } \right)^{1/2},\ \
\eta_{\rm psf} =   \left[ {\epsilon_{\bar{m}_I} t_{\rm exp} 
\over 4 n_{\rm div} }
\left( {\delta \theta_{\rm see} \over  \theta_{\rm see}} \right)^2
\right]^{1/2},
\eqno(6.1.2)
$$
where $\Sigma_{\rm sky}$ and $\Sigma_{\rm M31}$ are the surface 
brightnesses of the sky and M31, $\epsilon_{\bar{m}_I}$ is the rate of 
photon detection at the fluctuation magnitude, 
$\bar{m}_I  = 23.2\ {\rm mag}$ for M31 (Tonry, 1991), 
$\delta \theta_{\rm see} / \theta_{\rm see}$ is the fractional 
difference in seeing between images, and $n_{\rm div}$ 
is the number of sub-exposures in each exposure time $t_{\rm exp}$.
By dividing the total exposure time into smaller exposures, 
the noise due to the time variation of PSF decreases because $N_{\rm psf}$ 
is averaged out, while $N_{\gamma}$ remains the same.
As a result of the additional noise, the total noise, $N_{\rm tot} = 
(N_{\gamma}^2 + N_{\rm sky}^2  + N_{\rm psf}^2)^{1/2}$, 
is increased relative to the pure M31 photon noise by a factor,
$$
\eta = { N_{\rm tot} \over N_{\gamma} }
= (\eta_{\rm sky}^2 + \eta_{\rm psf}^2 )^{1/2}. 
\eqno(6.1.3)
$$
The net influence of additional noise on the event rate is that the 
required $(S/N)_{\rm min}$ increases by the factor $\eta$.

However, not all the events that satisfy the strict  $(S/N)_{\rm min}$ 
test qualify as detectable events.
Additional tests await for them.
One of these tests comes from the duration of an event: 
if the event lasts for very short time, the event could not be 
detected however high the $S/N$ is, because the event continues 
less than the observation interval (1 day).
This restriction of time scale is quantified in the computation by
counting only events with $t_{\rm eff} \ge 0.5\ {\rm day}$, 
where the effective duration time is defined by 
$ t_{\rm eff} = \beta_{\rm max} t_{\rm e}$ (Gould 1995). 
Some of these short events might occur during the 
observation which continues for $t_{\rm exp} = 4\ {\rm hrs}$
each night in average.
However, they would not be detected because of another 
restriction discussed below.

The other restriction comes from the finite size of 
the source star.
If the source size is bigger than the Einstein ring, 
$\theta_{\ast} > \theta_{\rm e}$, the point-source amplification 
formula in equation (1.2) is no longer valid at the central region of 
the light curve, i.e., when
$$ 
\beta_{\rm max} < {1 \over 2}{ \theta_{\ast} \over \theta_{\rm e} }
=  {1 \over 2} {t_{\ast} \over t_{\rm e} },
\eqno(6.1.4)
$$
where $t_{\ast} = (D_{\rm ol}/D_{\rm os})(R_{\ast} / v) 
\sim R_{\ast}/v$ is the crossing time over the star with a 
radius $R_{\ast}$.
Then the maximum amplification will be overestimated by equation (1.2) 
and the event with $0.5t_{\ast} > t_{\rm eff}$ cannot be 
detected (Gould 1995).
The luminosity at a certain frequency $\nu$ is 
$L_{\nu} = 4\pi^2 R_{\ast}^2 B_{\nu}$, where 
$B_{\nu} = (2 h\nu^3 /c^3)/[\exp (h\nu /kT) - 1]$ is the Planck function.
In the Rayleigh-Jeans limit, e.g., $K$ band 
$\lambda \sim 2.2\ {\rm \mu m}$, the Planck function is approximated by
$B_{\nu} \sim (2 \nu^2 / c^3) kT$, and thus the luminosity in the 
$K$ band is related to the stellar radius by
$R_{\ast} \propto (L_{K}/T)^{1/2}$.
Here the frequency is regarded constant over a narrow band.
Since the temperature of relevant source star varies within a 
narrow range, $3-6\times 10^3\ {\rm K}$, relative to the luminosity, 
$1 - 10^5\ L_{\odot}$, the dominant factor in determining 
stellar radius is the luminosity.
Therefore, one can approximate the stellar radius by
$R_{\ast} = 10^{0.2(M_{K,\odot}-M_{K,\ast})} R_{\odot}$, where 
the solar $K$-band luminosity is $M_{K,\odot} = 3.3\ {\rm mag}$ and 
$K$-band LF are provided from the model LF discussed in \S\ 6.2.

\subsection{Observational Strategy}

I begin by analyzing the event rate in a relatively modest 
observing program.
The experiment is carried out using a 
$2{\rm K} \times 2{\rm K}$ CCD camera with a pixel scale of
$0''\hskip-2pt.25$ on a 1 m telescope.
If the pixel scale is too big, the geometric alignment of the images 
would introduce considerable additional noise 
(pixelization error; Gould 1996).
I assume that the camera can detect $12\ {\rm photons}\ {\rm s}^{-1}$ 
for an $I = 20$ star.
The observation is carried out $4\ {\rm hrs}/{\rm night}$ 
during the M31 season ($\sim 1/3\ {\rm yr}$).
Bad weather will influence the event rate, e.g., by degrading the 
$S/N$ and by missing portions of the light curve.
However, the weather pattern is unpredictable and differs dramatically 
from site to site.
Hence, I set the observation time less than what might be being carried out 
($\sim 5-6\ {\rm hrs}/{\rm night}$) to compensate the loss of $\Gamma$ by the 
bad weather instead of modeling the weather pattern.
The total exposure time is subdivided into smaller integration times,
$\sim 20\ {\rm min}$, so that the images are not saturated.
The angular size of the PSF is
$\Omega_{\rm psf} \sim \pi \theta_{\rm see}^{2} / 
\ln 4 = 2.27\ {\rm arcsec}^2$, assuming typical seeing of
$\theta_{\rm see} \sim 1''$.
The fractional seeing difference is kept to 
$\langle \delta \theta_{\rm see} / \theta_{\rm see} \rangle \sim 0.05$
by using methods discussed in \S\ 7.
The surface brightness distribution is based on the density models of
the M31 bulge and disk and then normalized to the values of
Walterbos et al.\ (1987).
The LF is modeled primarily by adopting that of
Galactic bulge stars shown by dotted line in the second panel of Figure 3
(J.\ Frogel, private communication)

However, the LF in the faint end, $F < F_{\rm c}$, is incomplete.
Here $F_{\rm c}$ corresponds to $I=28.2$ where the LF begins to decline.
If the LF is adopted without corrections, the fraction of bright stars 
would be overestimated.
The net effect is then the systematic overestimation of event rate
because events are more likely to be detected for brighter stars.
Thus, I model the faint part of the LF ($F < F_{\rm c}$) by adopting 
the LF of stars in the
solar neighborhood determined by Wielen, Jahreiss, \& Kr\"uger
(1983) for $3.6 < M_I < 8$ and by
Gould, Bahcall, \& Flynn (1996) for $8 < M_I < 14$.
The corrected model LF yields the fluctuation magnitude
of $\sum_i \phi(F_{0,i})F_{0,i}^2 /\sum_i \phi(F_{0,i})F_{0,i}=23.48\
{\rm mag}$, which matches well with Tonry's (1991)
determination of $\bar{m}_{I}=24.3$ before the extinction correction.
The fraction of light from the stars in the corrected part of the LF
($M_I > 3.6$) is $14.7\%$.
The model LF is shown with a solid line in the second panel of Figure 3.

Combining all this information, the values of $\beta_{\rm max}$ are 
computed for a given stellar flux $F$ by solving equation (2.4) and 
taking the additional noise term $\eta$ into consideration.
The determined $\beta_{\rm max}$ as a function of the stellar flux is 
shown in the upper panel of Figure 3 
for the position $(x',y') \sim (1.0,1.0)\ {\rm kpc} $
at which $\bar{t}_{\rm e} \sim 16\ {\rm days}$, 
$\Sigma \sim 19\ {\rm mag}\ {\rm arcsec}^{-2}$, 
and $\tau \sim 4\times 10^{-6}$ for the various 
$(S/N)_{\rm min}$ noted in the figure.
The LF is normalized by the surface brightness 
of the position.
Note that as a star becomes fainter, it can only be detected 
at higher amplification and that the detection limit decreases as
$(S/N)_{\rm min}$ increases.
Now one is ready to compute event rate for any adopted $(S/N)_{\rm min}$
with the results of the $\tau$ and $t_{\rm e}$ distributions 
determined in \S\ 3 and \S\ 4.

\subsection{Event Rates}

The term ``event rate'' is somewhat subjective because its value depends 
on the required precision and strategy of the experiment.
To measure the time scale with better precision, higher $S/N$ is required.
The relation between the precision of measurement and the required $S/N$
is quantified by a factor 
$\Lambda = (S/N)\delta t_{\rm e}/ t_{\rm e}$, 
where $\delta t_{\rm e}/ t_{\rm e}$ is the fractional accuracy of the 
time scale measurement.
Its value is computed as a function of $\beta$ and shown in Figure 1 
of Gould (1996).
To determine $(S/N)_{\rm min}$, I use the mean value, 
$\langle \Lambda \rangle \sim 16$,
since $\Lambda$ is a slowly varying function of $\beta$ except in the 
region of very close encounter event (i.e., $\beta \leq 0.01$); 
$10 \leq \Lambda \leq 20$ in the range $0.1 \leq \beta \leq 0.8$.
The required minimum signal-to-noise ratio is then,  
$$
(S/N)_{\rm min} \sim { \langle \Lambda \rangle \over 
\delta t_{\rm e} / t_{\rm e} }.
\eqno(6.3.1)
$$

First I estimate the event detection rate based on the precision that 
can distinguish lensing events from flux variation due to statistical noise.
In principle it is possible to select events up to the Poisson 
limit of $7\sigma$ fluctuations, i.e., $(S/N)_{\rm min}=7$ (Gould 1996).
To be conservative, I use $(S/N)_{\rm min}=10$.
The resultant event rate map $d\Gamma (x',y')/d\Omega$, 
$d\Omega = 10^{3}\ {\rm arcsec}^{2}$, 
at this $(S/N)_{\rm min}$ 
is constructed and shown in Figure 4 as a contour map.
The contour levels are drawn in units of ${\rm events}\ {\rm yr}^{-1}$.
Then the total event rate detected over the entire 
area of the image, $\Omega_{\rm ccd} \sim 70\ {\rm arcmin}^2$, is computed by 
$$
\Gamma = \int_{\rm CCD} {d\Gamma \over d\Omega } d\Omega,
\eqno(6.3.2)
$$
resulting in $183\ {\rm events}\ {\rm yr}^{-1}$.

Now I stiffen the qualification of events by counting
only those that can provide time scales with an 
acceptable uncertainty, $\delta t_{\rm e} / t_{\rm e} \leq 0.2$. 
The required signal-to-noise ratio is then $(S/N)_{\rm min} = 80$ 
determined from equation (6.3.1).
The result is $\Gamma \sim 15\ {\rm events}\ {\rm yr}^{-1}$.
The total event rate for other values of $(S/N)_{\rm min}$ are 
shown in Figure 5 for individual lens population events.

\section{Improved Event Rate}

There are several ways of improving the event rate.
First, one can increase $\Gamma$ by reducing the seeing difference, 
$\delta \theta_{\rm see}$, which causes additional noise 
$\eta_{\rm psf}$.
The easiest way to reduce $\delta \theta_{\rm see} / \theta_{\rm see}$
is classifying images (there will be $\geq 1000\ {\rm yr}^{-1}$ of them) 
with similar seeing, and then subtracting the current image from 
reference images of similar seeing.
The other method, which is being developed by Tomaney \& Crotts
(1996), is reducing the seeing differences by
convolving the current image so that the resultant
PSF matches the reference frame.
Second, the event rate would increase by carrying out the experiment 
at more than one place.
By observing more often, one could detect events with shorter $t_{\rm eff}$
than could not be detected from a single site.
The other effect of multi-observation on $\Gamma$ is the increase 
of the time averaged photon detection rate, $\alpha$. 
Finally, a bigger telescope will detect more events.
The event rates with the additionally adopted observational strategies 
(single or combined) are shown in Table 1.
The values in the second and third columns 
represent $\Gamma$ when $(S/N)_{\rm min} =10 $ and 80, respectively.
The values in the parentheses represent the improved factors 
in $\Gamma$ compared to the value obtained under the observational 
condition described in \S\ 6.2.

If all of these strategies are adopted, the event rate would 
increase by up to a factor $\sim 6.5$, so one could obtain 
events with time scale data at a rate $\sim 62\ {\rm yr}^{-1}$.
Therefore, a few years of observation would provide one with a data 
set large enough for the analysis of M31 stars and MACHOs.
From this analysis, one can obtain important information such as 
the LF of M31 bulge stars (Gould 1996), the optical depth 
distribution of MACHOs, and above all the mass spectrum of M31 
MACHOs (Han \& Gould 1996a).

\section{Uncertainties}

The determined event rate is subject to many sources of uncertainties.
These include insufficiently detailed density and velocity distributions 
and mass spectrum of M31 bulge MACHOs. 
Additional uncertainties from the crude models of 
extinction and the LF also propagate to the 
determination of the event rate.

Among all the uncertainties, however, the most important comes 
from the stellar mass density and the fraction of dark matter 
in various forms, which  
can only be determined by the comparison of the theoretical 
predictions with the observational results.
The other uncertainties will not seriously affect the estimates 
given here.
For example, the velocity field is spectroscopically well constrained.
Although the model LF at the faint end is very uncertain, 
stars in this regime make little contribution to the event rate because
they are generally too faint to be detected. 
Thus, the detailed structure of the faint end of the LF 
is not important.
Of course, my determination of the amount of extinction, $f_V$, is
not precise enough for other type of studies such as the
construction of an extinction map which would require very 
detailed observation.
However, in the optically thin limit, the event rate depends 
only on the mean extinction (and even for a totally opaque 
disk only falls by a factor $\sim 1/2$), so the error induced by
uncertain extinction is unlikely to exceed 10\%.

In addition, it has been demonstrated by Tomaney \& Crotts (1996)
that systematic errors can be brought under very good control.
For example, they measure the flux variation of a variable 
found in the central M31 bulge (surface brightness 
$R\sim 18.2\ {\rm mag}\ {\rm arcsec}^{-2}$)
with a $1\sigma$ flux error equivalent to $R=24.6$, 
within a factor $\sim 2$ of the photon limit and 8 mag below the 
flux in a seeing disk.

\acknowledgements
I would like to thank A.\ Gould and J.\ Frogel  for
making very helpful comments and kindly providing the Galactic bulge 
luminosity function and to USNO for providing telescope time.
This work was supported by a grant AST 94-20746 from the NSF.

\newpage
\begin{table}
\begin{center}
\vskip3mm
\vskip3mm
\begin{tabular}{lcc}
\hline
\hline
\multicolumn{1}{c}{\ \ \ \ \ \ \ improved\ \ \ \ \ \ \ \ } &
\multicolumn{2}{c}{$\Gamma $} \\
\ \ \ \ \ \ \ \ strategies & $\ \ \ (S/N)_{\rm min} = 10\ \ \ \ \ \ \ $ 
& $\ \ \ (S/N)_{\rm min} = 80\ \ \ \ \ \ \ $
\\
\hline
$\delta\theta_{\rm see} / \theta_{\rm see}=0.05$ & 366\ (2.0) & 43.7\ (2.8) \\
\bigskip
2-m telescope &  &  \\
$\delta\theta_{\rm see} / \theta_{\rm see}=0.05$ & 293\ (1.6) & 31.2\ (2.0)  \\
\bigskip
three-site observation &   &  \\
\bigskip
$\delta\theta_{\rm see} / \theta_{\rm see}=0.01$ & 201\ (1.1) & 18.7\ (1.2) \\
$\delta\theta_{\rm see} / \theta_{\rm see}=0.01$ & 384\ (2.1) & 43.7\ (2.8)  \\
\bigskip
three-site  observation  &   &   \\
$\delta\theta_{\rm see} / \theta_{\rm see}=0.01$ & 403\ (2.2) & 48.3\ (3.1) \\
\bigskip
2-m telescope &   &  \\
$\delta\theta_{\rm see} / \theta_{\rm see}=0.01$ & 732\ (4.0) & 101.4\ (6.5) \\
three-site  observation  &  &  \\
2-m telescope  & &  \\
\hline
\end{tabular}
\caption{
The event rates with improved observational strategies in units of
${\rm events}\ {\rm yr}^{-1}$.
The values in the parentheses represent the improved factors
compared to that under observational condition described in \S\ 6.2.
The first column shows the additionally adopted strategies,
and the values in the second and third columns are those of $\Gamma$ when
$(S/N)_{\rm min} = 10$ and 80, respectively.}
\end{center}
\end{table}

\newpage
\centerline{\bf FIGURE CAPTION}
\bigskip
 
\noindent
{\bf Figure 1}:
The resulting optical depth distribution caused by M31 population
(M31 disk+bulge self-lensing plus M31 halo) is shown as a 
function of projected position $(x',y')$.
The total optical depth both by M31 and Galactic halo lenses are
obtained by adding $\tau_{\rm MW,halo}=4.4\times 10^{-7}$ to the values marked
on the contour.
The contour levels are marked on the map in units of $10^{-6}$.
The lower half of the figure contains the near side of the M31 disk.

\noindent
{\bf Figure 2}:
Median time scale distribution as a function of
$(x',y')$ for M31 self-lensing events.  
The contour levels are drawn in units of ${\rm days}$.
Other population events, Galactic halo and M31 halo events, the time scale 
distributions are similar to that of M31 self-lensing events.
 
\noindent 
{\bf Figure 3}:
Values of $\beta_{\rm max}$ as a function of
luminosity of M31 bulge stars (upper panel) and
the adopted LF of M31 bulge stars (lower panel).
For the computation of $\beta_{\rm max}$, values of $\Sigma$ and
$\bar{\omega}$ at $(x',y')=(1.0,1.0)\ {\rm kpc}$ are used.
The basic LF function (dotted line) is modeled based on the 
Galactic bulge LF (J.\ Frogel, private communication).
The faint end of the LF (solid line) is modeled by combining
the LFs of Gould et al.\ (1996) and Wielen et al.\ (1983).
 
\noindent 
{\bf Figure 4}:
The event rate per angular area, $d\Omega = 10^3\ {\rm arcsec}^2$,
as a function of position $(x',y')$
for the threshold signal-to-noise ratio of $(S/N)_{\rm min}=10$.
The contour levels are drawn in units of ${\rm events}\ {\rm yr}^{-1}$.
 
\noindent 
{\bf Figure 5}:
The total event rate, $\Gamma =(d\Gamma / d\Omega )\Omega_{\rm ccd}$,
for individual population events
as a function of $(S/N)_{\rm min}$.

\begin{figure}
\postscript{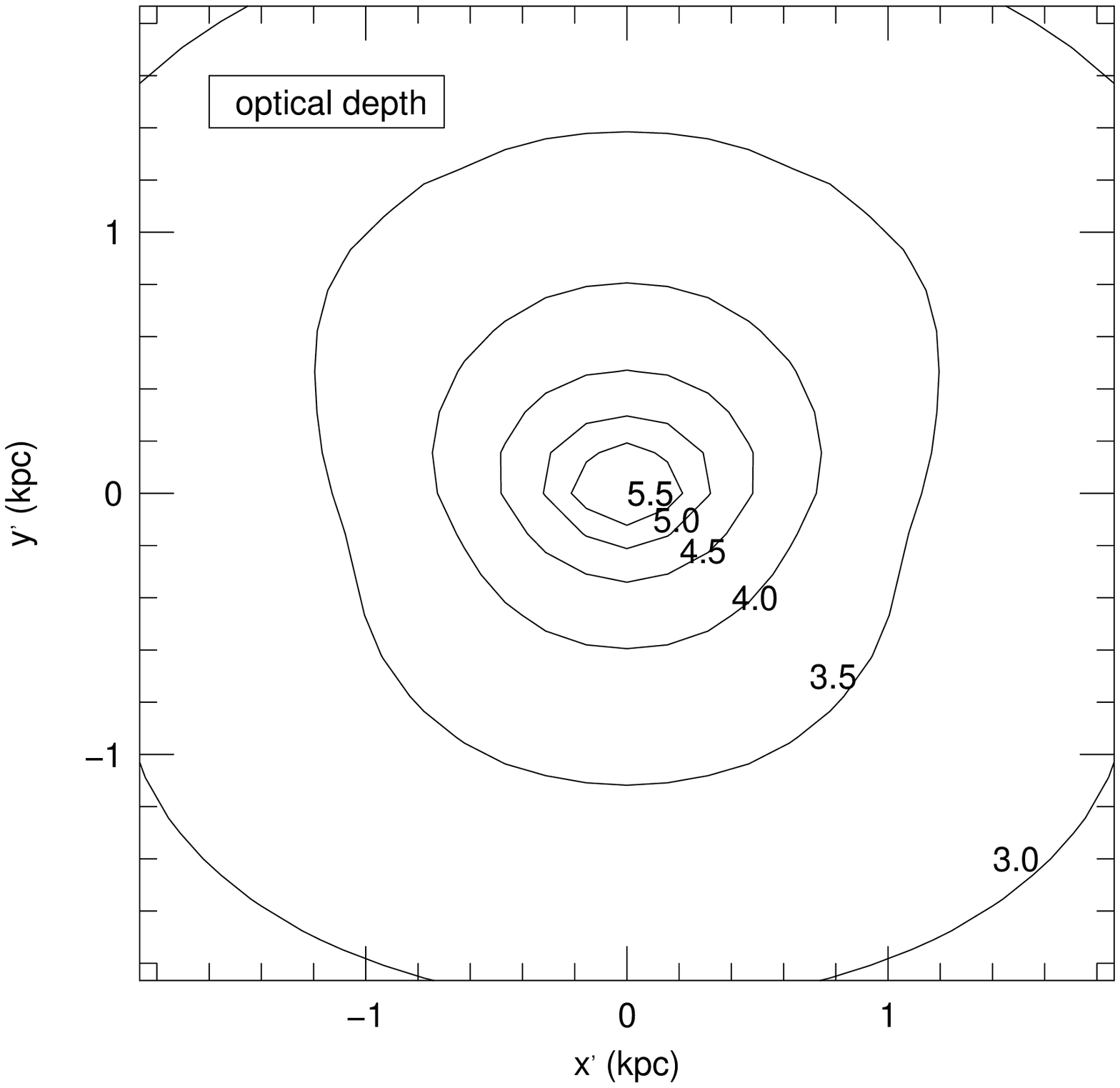}{1.1}
\end{figure}

\begin{figure}
\postscript{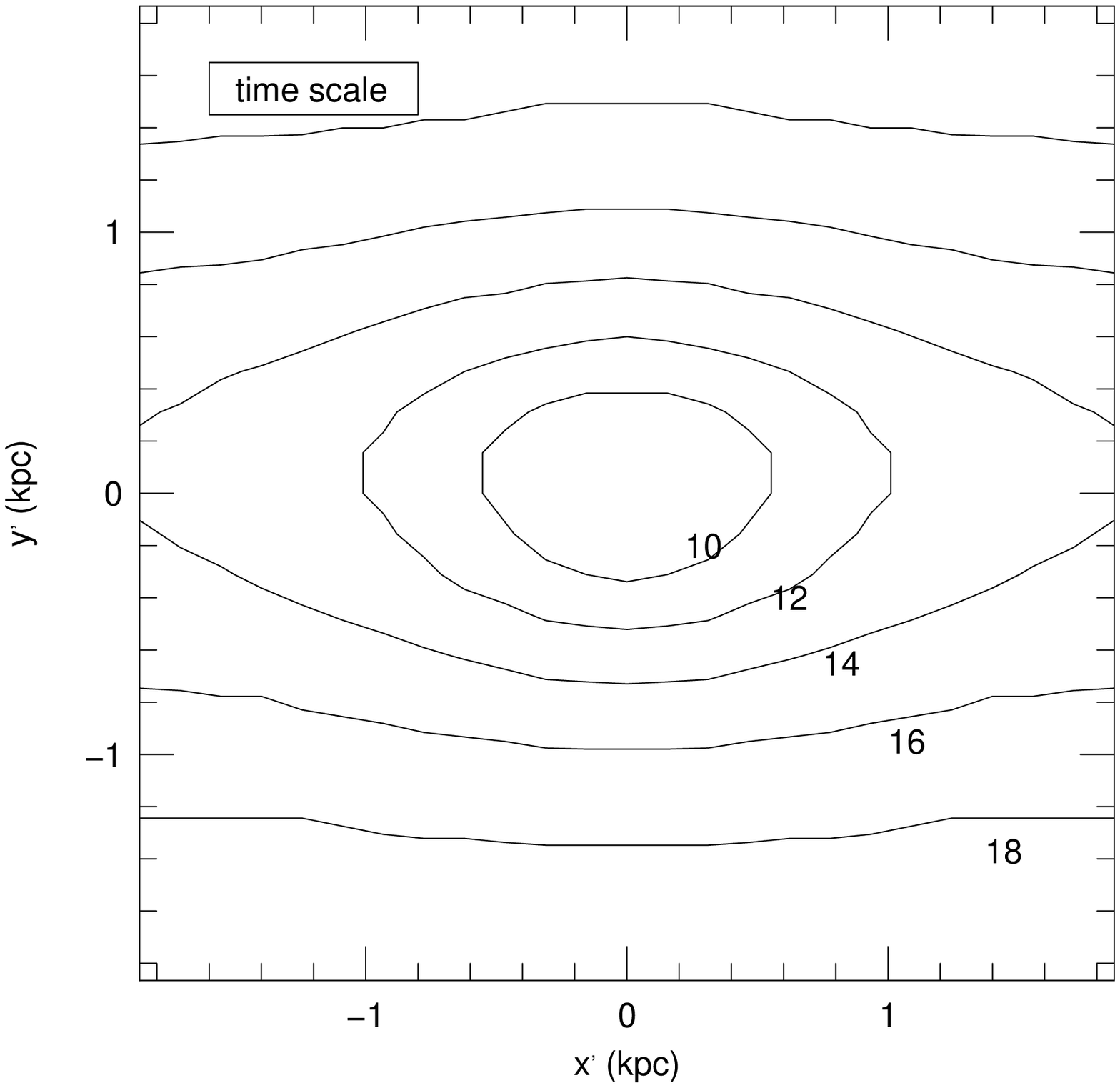}{1.1}
\end{figure}

\begin{figure}
\postscript{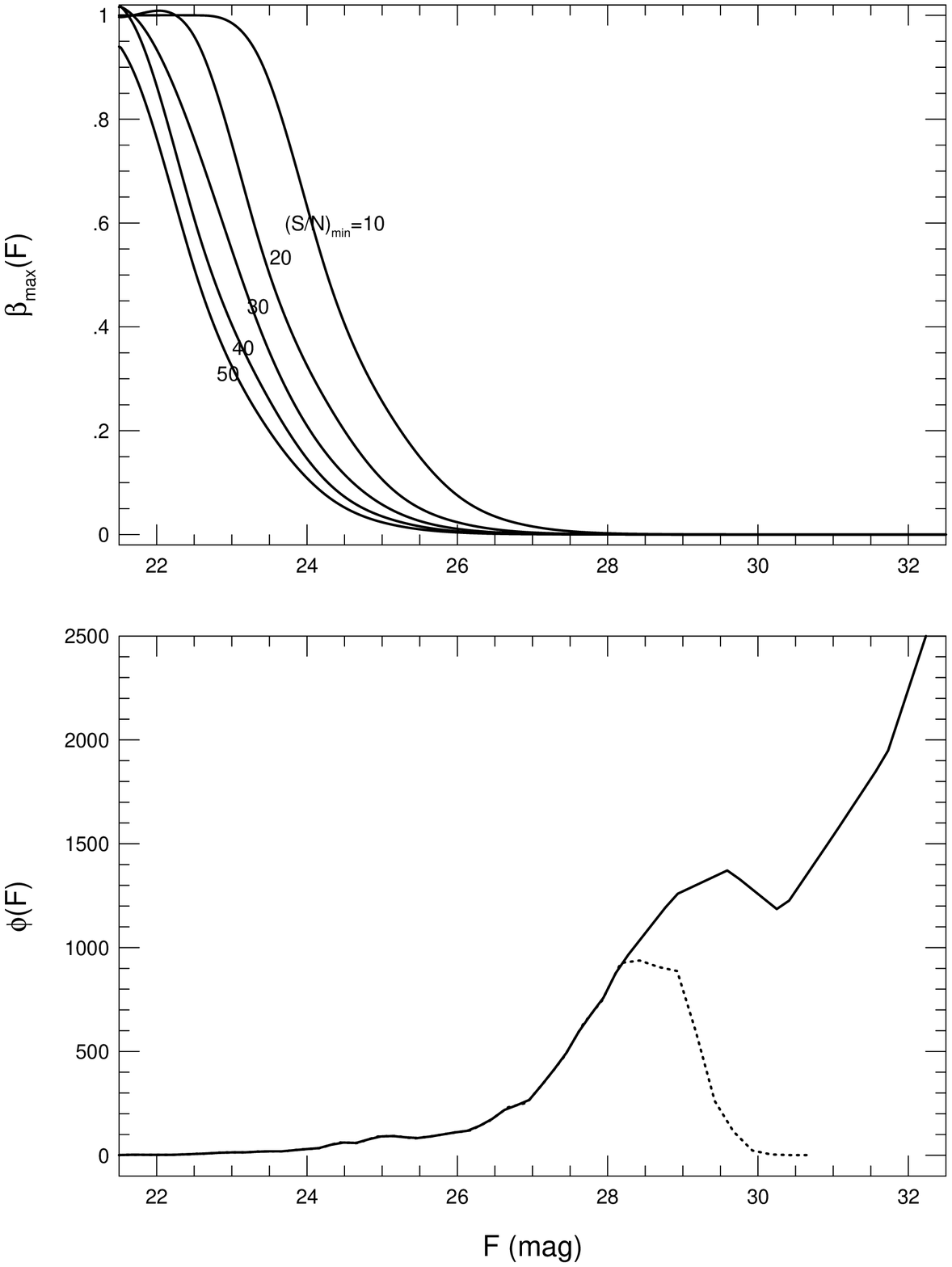}{1.1}
\end{figure}

\begin{figure}
\postscript{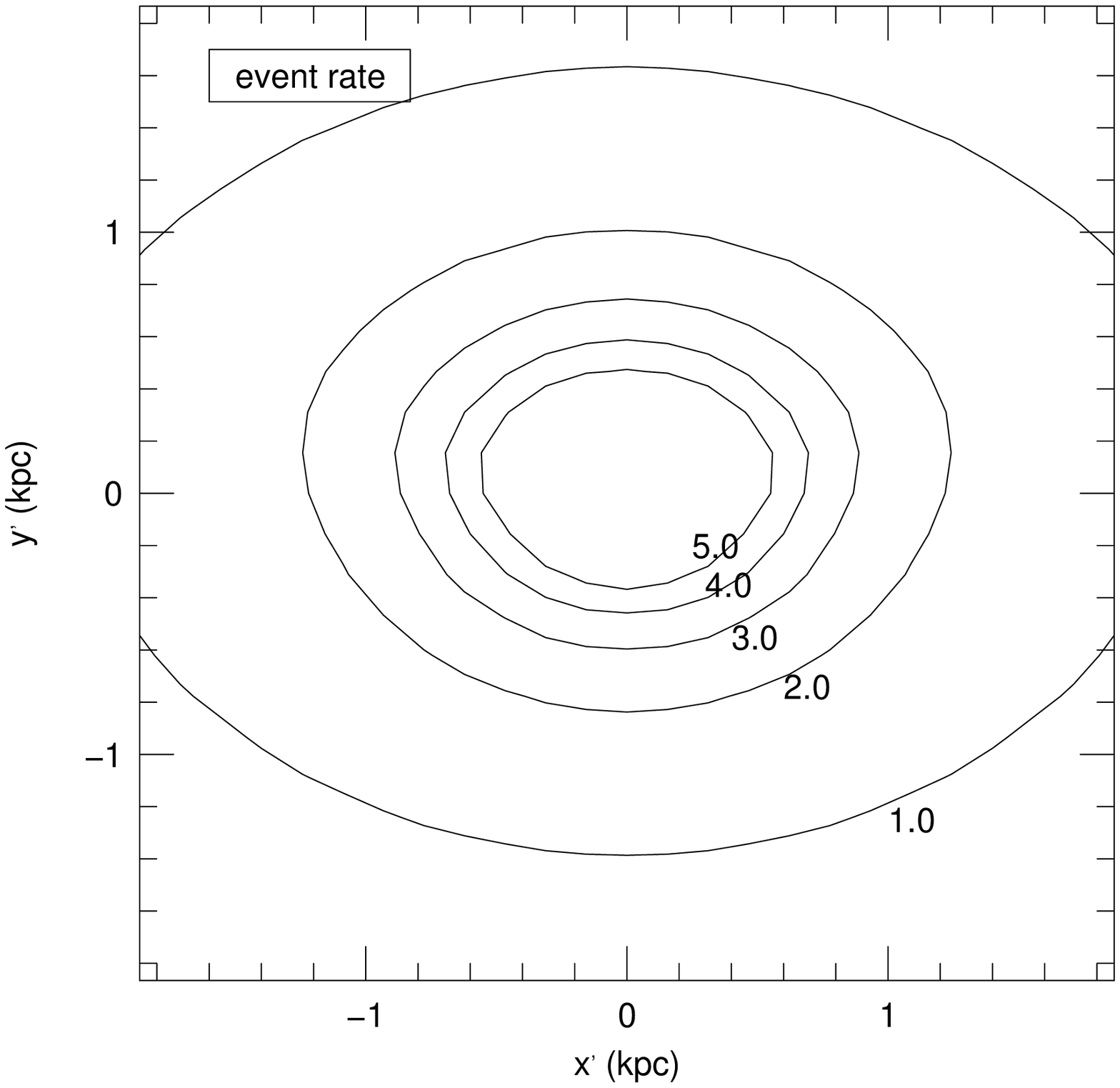}{1.1}
\end{figure}

\begin{figure}
\postscript{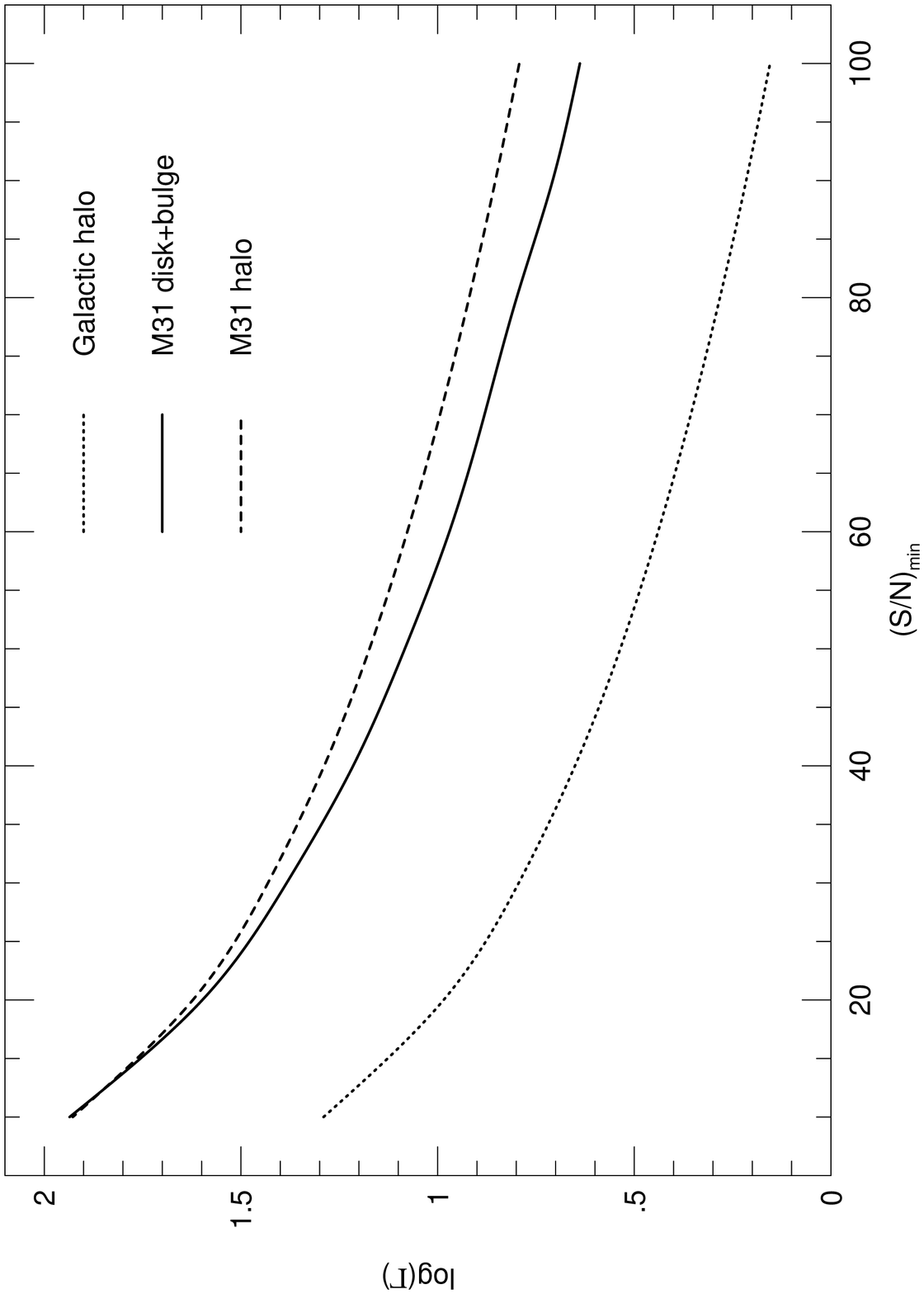}{1.1}
\end{figure}


\newpage
\begin{references}

\reference Alard, C., Guibert, J., Bienayme, O., Valls-Gabaud, D., 
           Robin, A.\ C., Terzan, A., \& Bertin, E.\ 1995, Msngr, 80, 31
\reference Alcock, C.\ et al.\ 1995, PhRvL, 74, 2867
\reference Ansari, R.\ et al.\ 1995, A\&A, 299, L21
\reference Ansari, R.\ et al.\ 1996, 2nd Workshop of 
           `` The Dark Side of the Universe: experimental efforts 
           and theoretical frame work''
\reference Bahcall, J.\ A., Schmidt, M., \& Soneira, R.\ M.\ 1983,
	   ApJ, 265, 730
\reference Baillon, P., Bouquet, A., Giraud-H\'eraud, Y., \& Kaplan, J.\ 
           1993, A\&A, 277, 1
\reference Braun, B.\ 1991, ApJ, 372, 54
\reference Colley, W.\ N.\ 1995, AJ, 109, 440
\reference Crotts, A.\ P.\ S.\ 1992, ApJ, 399, L43
\reference Gilmore, G., King, I.\ R., \& van der Kruit, P.\ C.\ 1990, 
	   The Milky Way as a Galaxy (University Science Book, 
           Mill Valley), 35
\reference Gould, A.\ 1994, ApJ, 435, 573
\reference Gould, A.\ 1995, ApJ, 455, 44
\reference Gould, A.\ 1996, ApJ, 470, 000
\reference Gould, A., Bahcall, J.\ A., \& Flynn, C.\ 1996, ApJ, 465, 000
\reference Han, C., \& Gould, A.\ 1996, ApJ, 467, 000
\reference Han, C., \& Gould, A.\ 1996b, ApJ, submitted
\reference Kent, S.\ 1989a, PASP, 101, 489
\reference Kent, S.\ 1989b, AJ, 97, 1614
\reference Lawrie, D.\ G.\ 1983, ApJ, 273, 562
\reference Tomaney, A.\ B., \& Crotts, A.\ P.\ S.\ 1996, ApJ, submitted
\reference Tonry, J.\ L.\ 1991, ApJ, 373, L1
\reference Udalski, A.\ et al.\ 1995, AcA, 45, 237
\reference Walterbos, R.\ A.\ M.\ \& Kennicutt, Jr.\ R.\ C.\ 1987, 
           A\&AS, 69, 311
\reference Wielen, R., Jahreiss, H., \& Kr\"uger, R.\ 1983, IAU coll.\ 76:
	   Nearby Stars and the Stellar Luminosity Function, eds.\ 
	   A.\ G.\ D.\ Philip \& A.\ R.\ Upgren, p.163 
\reference Zhao, H., Spergel, D.\ N., \& Rich, R.\ M.\ 1995, ApJ, 440, L13
 
\end{references}
\end{document}